\documentclass[twocolumn,pra]{revtex4-1}

\usepackage{amssymb, amsmath, amsthm}	
\usepackage{graphicx}	
\usepackage{color}
\usepackage[colorlinks=true,linkcolor=red,citecolor=blue,urlcolor=green]{hyperref}
\usepackage{units}

\newcommand{\ket}[1]{\ensuremath{\left|{#1}\right>}}
\newcommand{\subt}[2]{{#1}_{\text{{#2}}}}

\begin{document}

\title{Effect of buffer gas on electromagnetically induced transparency in a ladder system using thermal rubidium vapor}

\author{Armen Sargsyan}
\author{David Sarkisyan}
\affiliation{Institute for Physical Research, National Academy of Sciences - Ashtarak 2, Armenia}

\author{Ulrich Krohn}
\email[Electronic address: ]{u.m.krohn@durham.ac.uk}
\author{James Keaveney}
\author{Charles Adams}

\affiliation{Department of Physics, Rochester Building, Durham University, South Road, Durham DH1 3LE, United Kingdom}

\date{\today}

\begin{abstract}
We report on the first observation of electromagnetically induced transparency (EIT) in a ladder system in the presence of a buffer gas.
In particular we study the 5S$_{1/2}$ - 5P$_{3/2}$ - 5D$_{5/2}$ transition in thermal rubidium vapor with a neon buffer gas at a pressure of 6~Torr. In contrast to the line narrowing effect of buffer gas on $\Lambda$--systems we show that the presence of the buffer gas leads to an additional broadening of $(32 \pm 5)$~MHz, which suggests a cross section for Rb(5D$_{5/2}$)--Ne of $\sigma_{\text{k}}^{\text{(d)}} = (7\pm 1)\times 10^{-19}\,\text{m}^2$. However, in the limit where the coupling Rabi frequency is larger than the collisional dephasing a strong transparency feature can still be observed.
\end{abstract}

\maketitle

The effect of electromagnetically induced transparency (EIT) arises due to coherence in three-level systems as first described in \cite{Arimondo:1976p1393, Harris:1990p1159}, and experimentally demonstrated in \cite{Boller:1991p1316}. Typically, the three-level system consists of two long-lived states (\ket{1} and \ket{3}), which are coupled by two lasers, labelled probe and coupling with Rabi frequencies $\subt{\Omega}{p}$ and $\subt{\Omega}{c} (> \subt{\Omega}{p})$, to a radiative state \ket{2} with a lifetime, $1/\Gamma_2$. 
If the two lasers are resonant, i.e., the detunings $\delta_{12} = \delta_{23} = 0$, 
the imaginary part of the one-photon coherence 
$\text{Im}(\rho_{12})$ and the absorption coefficient $\alpha \propto \text{Im}(\rho_{12})$ are zero, rendering the medium fully transparent \cite{Fleischhauer:2005p1240}. For $\Omega_{\rm p}<\Gamma_2$ the transparency is caused by a destructive interference of the excitation amplitudes into the intermediate state \ket{2}, which results in the occupation of a dark state $\ket{\emptyset}\simeq\ket{1}$ with no contribution from the radiative state. The width of the transparency window is determined by the dephasing rate between states \ket{1} and \ket{3}, $\Gamma_{13}$, and can be much narrower than the natural linewidth, $\Gamma_{13}<\Gamma_2$. For photon storage applications one is interested in reducing the dephasing rate as the narrow resonance results in a large group index \cite{Liu:2001p1507} and enables long photon storage times \cite{Schnorrberger:2009p1508}. For thermal atoms, the dephasing rate can be reduced by using the hyperfine ground states as the long lived states forming a $\Lambda$--system, and a buffer gas to increase the interaction time with the laser beams \cite{Brandt:1997p1541, Lukin:1997p1152, Javan:2002p1398}. 

Another topic of interest is the ladder or cascade system where \ket{3} is a higher energy excited state \cite{GeaBanacloche:1995p576, Shepherd:1996p1399, Boon:1999p1501}. For example, if state \ket{3} is a Rydberg state \cite{Mohapatra:2007p20}, this opens interesting possibilities for quantum information \cite{Friedler:2005p1542, Muller:2009p815, Low:2009p197} and electrometry \cite{Mohapatra:2008p1510, Bason:2008p38}.  In the ladder system the dephasing rate is typically larger than for $\Lambda$--systems due to the spontaneous decay of level \ket{3}. Otherwise, EIT in $\Lambda$ and ladder systems show generally similar properties. However, the addition of a buffer gas changes this behavior dramatically. 
In $\Lambda$--systems the dephasing of the ground state coherences due to the collisions with the buffer gas is negligible and, instead the increased transient time due to collisional diffusion allows an extremely narrow linewidth to be observed \cite{Brandt:1997p1541}. In contrast, for cascade systems the collisional dephasing due to the buffer gas becomes the dominant line broadening mechanism and it is predicted that the EIT threshold (coupling power required to observed a transparency) and linewidth increases monotonically with the buffer gas pressure \cite{Kumar:2009p1158}. However, as suggested in \cite{Kumar:2009p1158} one still expects to  see a transparency effect if the Rabi frequency of the coupling laser is larger than the dephasing rate. In this regime, where $\subt{\Omega}{c} > \Gamma_2$, the medium is rendered transparent by means of Autler-Townes splitting of the resonance and the transparency effect is no longer dependent on destructive interference. It should be noted that below the terms transparency and EIT are used in a broader sense, and not restricted to coherent excitation of a dark state. To our knowledge EIT in a ladder system in the presence of a buffer gas has not been observed previously. However, it remains generally interesting to study such transparency features in the presence of large dephasings as for example found for atom-surface interactions in sub-micron cells \cite{Fichet:2007p1149, Kubler:2010p1598}.

\begin{figure}[b]
 \includegraphics[width=0.4\textwidth,angle=0]{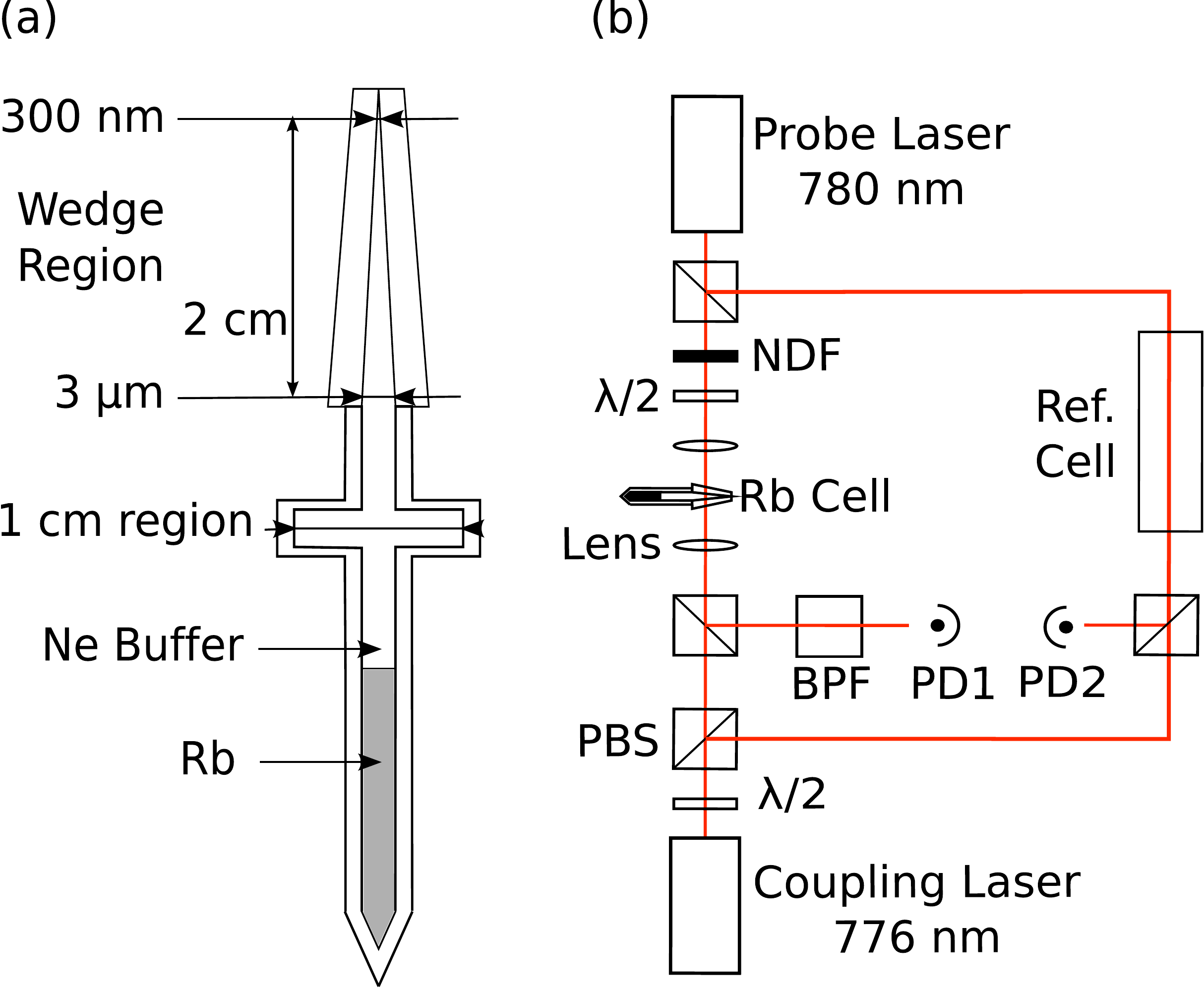}
\caption{(a) Schematic of the Rb vapor cell. The cell consists of two separate regions - a region with thickness of \unit[10]{mm} and wedge, where the thickness varies between \unit[300]{nm} and \unit[3]{$\mu$m} over a region of approximately \unit[2]{cm}.
 (b) Schematic of the experimental setup. A reference cell of length \unit[10]{cm} is used for frequency calibration. A photodiode PD1 (PD2) is used to record the signal from the buffer cell (reference cell). NDF - Neutral Density Filter; PBS - polarising beam splitter;  BPF - band pass filter; $\lambda/2$ - half wave-plate.}
 \label{fig:1}
\end{figure}

In this paper we demonstrate ladder EIT in Rb thermal vapor in the presence of a buffer gas. We show that collisions with the buffer gas lead to a broadening of the transparency window. We show that for the 5S-5P-5D system in Rb vapor and Ne buffer gas (\unit[6]{Torr}) leads to an additional broadening of $\unit[(32 \pm 5)]{MHz}$, which is consistent with a Rb(5D)--Ne cross section of $\subt{\sigma}{k}^{(\text{d})}=\unit[(7\pm 1)\times 10^{-19}]{m^2}$.

A schematic of the experimental setup is shown in figure \ref{fig:1}. Probe and coupling laser beams are produced by extended cavity diode lasers operating at wavelengths $\subt{\lambda}{p}=\unit[780]{nm}$ and $\subt{\lambda}{c}=\unit[776]{nm}$, respectively. The beams counterpropagate through a Rb cell to minimize the Doppler broadening of the transition to the residual Doppler width $\delta\subt{\omega}{d} = |\subt{\omega}{p} - \subt{\omega}{c}|\,v/c \simeq \unit[2\pi\times 2]{MHz}$ for a temperature of $T=\unit[157]{^\circ C}$. Lenses with a focal length of \unit[20]{cm} are used to create a spot size ($1/\text{e}^2$ diameter) in the cell of $d=\unit[27]{\mu m}$. The time of flight of the atoms through the beam is $\tau=d/v\simeq \unit[80]{ns}$ in a thermal rubidium vapor and, hence, a transient time broadening in the absence of collisions with the buffer gas $\delta\omega_{\tau} = \tau^{-1} \simeq\unit[2\pi\times 2]{MHz}$. In the presence of the Ne buffer gas the transient time broadening is reduced by almost three orders of magnitude \cite{Wynands:1999p1599}.
 
In order to calibrate the frequency axis a part of each beam is directed through a reference Rb vapor cell of \unit[10]{cm}  length. The coupling laser is resonant with the 5P$_{3/2} \rightarrow$ 5D$_{5/2}$ ($\ket{2}\rightarrow\ket{3}$) transition, i.e., $\subt{\delta}{c} = 0$. The probe laser is scanned across the frequency range of the D2 line ($\ket{1}\rightarrow\ket{2}$). Both laser beams have orthogonal linear polarization. A narrow band pass filter (Semrock LL01-780-12.5) has been mounted in front of PD1 to remove residual \unit[776]{nm} light on the detector. 
\begin{figure}[b]
\includegraphics[width=0.48\textwidth,angle=0]{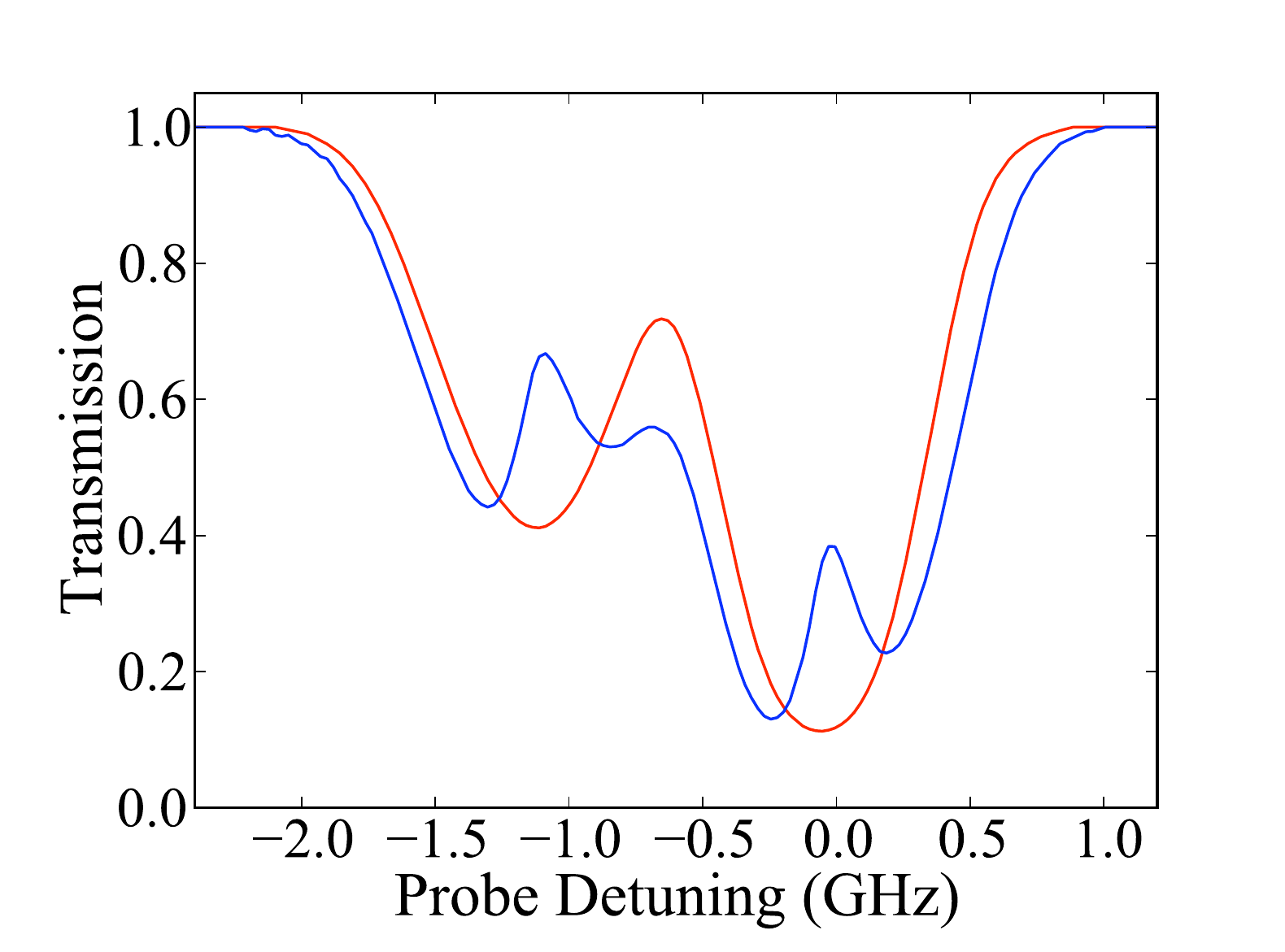}
 \caption{Transmission spectra as a function of the probe detuning through the \unit[10]{mm} buffer gas cell, measured at T = \unit[$(45 \pm 1)$]{$^\circ$C} without the coupling beam (red line). The power in the probe beam was $\subt{P}{p} = \unit[1]{\mu W}$. The zero of the detuning scale is set to the frequency of the $^{85}$Rb 5S$_{1/2}\, \subt{F}{g} = 3 \rightarrow$ 5P$_{3/2}\,\subt{F}{e} = 4$ transition. The left spectral line corresponds to the $^{87}$Rb 5S$_{1/2}\, \subt{F}{g} = 2 \rightarrow$ 5P$_{3/2}\,\subt{F}{e} = {1,2,3}$ transition. Switching on the coupling beam (blue line) with a power of $\subt{P}{c} = \unit[196]{mW}$ results in the coupling of $\subt{F}{e}$ to the 5D$_{5/2}$ manifold.}
\label{fig:2}
\end{figure}

The experiments are performed in two cells of similar construction, one pure rubidium cell, and a buffer gas cell with the addition of 6 Torr of Ne. The ground-state Rb--Ne collision rate is estimated to be \cite{Erhard:2001p1593}
\begin{gather}\label{eq:1}
 \subt{R}{k}^{\text{(s)}} = \frac{p}{\subt{k}{B}T}\,\subt{\sigma}{k}^{(\text{s})}\bar{v} \simeq \unit[17]{MHz}
\end{gather}
The Rb--Ne collisions lead to a narrowing of the EIT linewidth in a $\Lambda$--system as the reduction in the transit time broadening dominates over collisional dephasing within the ground state hyperfine structure. However, in a cascade system the collisional dephasing between ground and excited states cannot be neglected and leads to a broadening of the EIT linewidth.
A typical spectrum of the Doppler broadened $^{85}$Rb and $^{87}$Rb transition at a temperature of \unit[(45 $\pm$ 1)]{$^{\circ}$C} in the \unit[10]{mm} buffer gas cell can be seen in figure \ref{fig:2}. Distinctive transparency peaks appear when the coupling laser is switched on and when the coupling Rabi frequency is sufficiently strong to overcome the collisional dephasing. For these parameters the width of the transparency peaks is \unit[(105 $\pm$ 5)]{MHz} obtained by fitting a Gaussian function to the profiles. Note that no transparency was observed for a cascade system involving highly excited Rydberg states as the coupling laser power was insufficient to fulfill the condition $\Omega_{\rm c} > \Gamma_{13}$.

\begin{figure}[ht]
\includegraphics{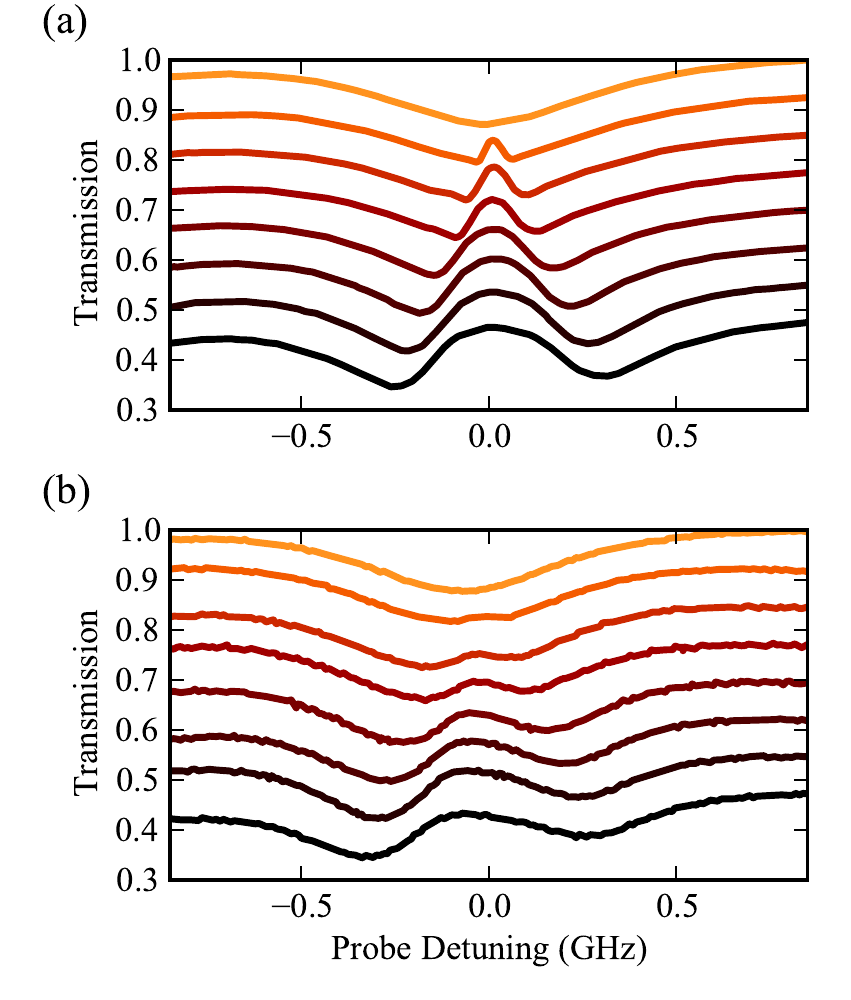}
 \caption{Transmission of the $^{85}$Rb $\subt{F}{g} = 3 \rightarrow \subt{F}{e} = 2,3,4$ transitions through the $\sim\unit[780]{nm}$ thick region of the cell, (a) without and (b) with \unit[6]{Torr} Ne buffer gas, for coupling laser powers of (top to bottom) 0, \unit[4]{mW}, \unit[14]{mW}, \unit[28]{mW}, \unit[70]{mW}, \unit[112]{mW}, \unit[154]{mW} and \unit[196]{mW}. The power of the probe laser was \unit[1]{$\mu$W}, and the temperature of the cell was $\unit[(157 \pm 10)]{^{\circ}C}$. An offset in transmission of 0.075 has been added for clarity to all spectra except the topmost.}
\label{fig:3}
\end{figure}

For further investigation of the broadening of the transparency feature due to the buffer gas we focus on the $^{85}$Rb feature. The experiments were performed in the wedge region shown in figure \ref{fig:1}, with the thickness of the atomic sample of order $\sim\unit[780]{nm}$ to allow tight focussing of the laser beams while still ensuring uniform beam size along the propagation direction. The mean free path of the atoms in the \unit[780]{nm} region is larger than the beam size, such that no additional narrowing due to collisions is observed. Sargsyan \emph{et al.} report on experiments with thermal atomic samples of the same size without buffer gas background exploiting a $\Lambda$--system \cite{Sargsyan:2006p1600}.

\begin{figure}[ht]
 \includegraphics{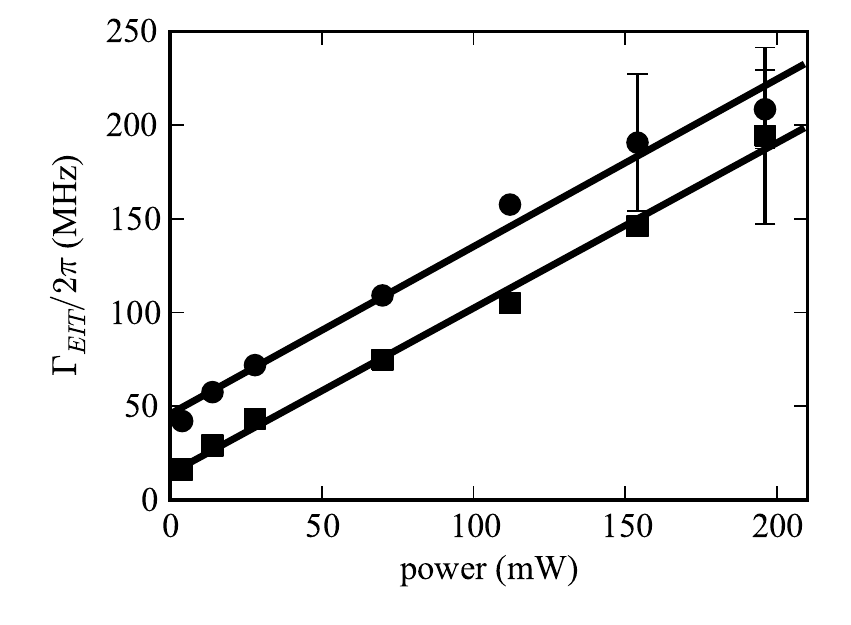}
 \caption{Width of the EIT feature against coupling laser power for the $^{85}$Rb 5S$_{1/2}\,\subt{F}{g} = 3 \rightarrow$ 5P$_{3/2}\,\subt{F}{e} = {2,3,4} \rightarrow$ 5D$_{5/2}$ transition without ($\blacksquare$) and with buffer gas ({\large$\bullet$}), respectively. The widths were obtained by a fit with a Gaussian function to the data shown in Fig. \ref{fig:3}. The thickness of the atomic sample is $\sim\unit[780]{nm}$. The width shows a linear dependence on the coupling laser power with a slope of $\unit[(893\pm 49)]{MHz/W}$. Adding \unit[6]{Torr} of Ne buffer gas to the atomic sample results in an additional broadening of $\unit[(32\pm 5)]{MHz}$. Note that the error bars, given by the standard deviation, increase as the Gaussian fit function becomes inaccurate in the Autler-Townes limit.}
\label{fig:4}
\end{figure}

The spectra in figure \ref{fig:3} show the same Autler-Townes splitting of the Doppler broadened resonance as in figure \ref{fig:2} both without (a) and with (b) the buffer gas. The additional broadening due to the buffer gas is clearly observed. The width of the transparency feature $\subt{\Gamma}{EIT}$ is obtained by fitting a function of the form $\mathcal{T}\sim 1 - \subt{\mathcal{A}}{d}\text{exp}(-(\Delta^2/(2\delta\subt{\omega}{d}^2)) + \subt{\mathcal{A}}{EIT}\text{exp}(-(\Delta^2/(2\subt{\Gamma}{EIT}^2))$ to the transmission spectra $\mathcal{T}(\Delta)$.
Figure \ref{fig:4} shows the dependence of the transparency linewidth versus the Rabi frequency of the the coupling laser,  $\subt{\Omega}{c}$. 
This function allows to quantify the transparency linewidth for low powers ($\subt{P}{c}\lesssim\unit[150]{mW}$), but shows an increasing deviation for larger powers, as indicated by the error bars in figure \ref{fig:4}.

The width as a function of the coupling power $\subt{P}{c}\sim \subt{\Omega}{c}^2$ shows a linear dependence as expected from \cite{Erhard:2001p1593}:
\begin{gather}
  \label{eq:Gamma_EIT}
 \subt{\Gamma}{EIT} = \Gamma_{11} + \Gamma_{13} + \Gamma_{31} + \Gamma_{33} + \subt{\Delta}{EIT}^{\text{c}} + \frac{\subt{\Omega}{c}^2}{\gamma + \subt{\delta}{c}^2/\gamma}
 \intertext{with}
 \gamma = \frac{1}{2}(\gamma_{21} + \gamma_{23} + \gamma_{2\text{r}} + 2\Delta^{\text{c}})
\intertext{and}
\Delta^{\text{c}} = \Delta^{\text{c}}_{11} + \Delta^{\text{c}}_{22} - 2 \Delta^{\text{c}}_{12} = \Delta^{\text{c}}_{33} + \Delta^{\text{c}}_{22} -2 \Delta^{\text{c}}_{32}
\end{gather}
where the $\Gamma_{ij}$ (collisions) and $\gamma_{ij}$ (radiative decay) are the incoherent population transfer rates between the states \ket{i} to \ket{j} and $\Delta_{ij}^{\text{c}}$ are the collisional induced decay rates of the respective coherences. State \ket{\text{r}} is a reservoir, which considers decay out of the ladder system. \\
The collisional broadening $\subt{\Delta}{EIT}^{\text{c}}$ due to collisions between Rb atoms and the buffer gas are
\begin{gather}
 \subt{\Delta}{EIT}^{\text{c}} = 2(\Delta^{\text{c}}_{11}+\Delta^{\text{c}}_{33}-2 \Delta^{\text{c}}_{13})
\end{gather}

The linear dependence of the width of the transparency resonance on the coupling laser power shows a slope of $\gamma^{-1} \propto 2\pi\times\unit[(893\pm 49)]{MHz/W}$ and an additional broadening of $\subt{\Delta}{EIT}^{\text{c}} = 2\pi\times\unit[(32\pm 5)]{MHz}$ due to the presence of the \unit[6]{Torr} Ne buffer gas. Since the slope $\gamma^{-1}$ in equation \eqref{eq:Gamma_EIT} is equal within the error bars for the experiments with and without the buffer gas we conclude that the differential collisional induced decoherence rate between the ground state 5S$_{1/2}$ and the intermediate state 5P$_{3/2}$ and between 5P$_{3/2}$ and 5D$_{5/2}$ is $\Delta^{\text{c}}\simeq 0$, or at least small in comparison to the differential broadening $\subt{\Delta}{EIT}^{\text{c}}$ between the 5S$_{1/2}$ and the 5D$_{5/2}$ state. Assuming equation \eqref{eq:1} the value for the collisional cross section between the 5D$_{5/2}$ state and the neon buffer gas is $\subt{\sigma}{k}^{\text{(d)}} = \unit[(7\pm 1)\times 10^{-19}]{m^2}$, compared to $\subt{\sigma}{k}^{\text{(s)}} \approx \unit[4\times 10^{-19}]{m^2}$ for the ground state \cite{Erhard:2001p1593}.

In conclusion we have for the first time experimentally shown that a transparency resonance may be observed in a cascade level scheme in the presence of a buffer gas as long as the coupling field is sufficient strong, such that the Autler-Townes splitting is larger than the collisional dephasing. In future experiments it might be interesting to study the dependence of the EIT linewidth on the buffer gas pressure in this ladder system in order to get an accurate value for the collisional cross section $\subt{\sigma}{k}^{\text{(d)}}$. 

Although for a ladder system the addition of buffer gas results in increased broadening, in some experimental situations the Rb-Ne collisions might be advantageous. In particular, when the laser spot sizes become small ($\approx\unit[1]{\mu m}$) the transient time broadening in a pure atomic vapor ($\approx\unit[50]{MHz}$) exceeds the collisional broadening between the atoms and the buffer gas. In this case the addition of a buffer gas with a low pressure ($\approx\unit[1]{Torr}$) would result in the reduction of the transient time broadening while leading to only a small additional collisional broadening ($\approx\unit[10]{MHz}$).

We acknowledge the financial support from the Engineering and Physical Sciences Research Council (EPSRC).

\end{document}